\documentclass[10pt,twoside]{amundi_article}

\usepackage[table]{xcolor}
\usepackage{amssymb}
\usepackage{amsfonts}
\usepackage{amsmath}
\usepackage{amsthm}
\usepackage{fancybox}
\usepackage{fancyhdr}
\usepackage{graphicx}
\usepackage{subcaption}
\usepackage[hyphens]{url}
\usepackage{arydshln}
\usepackage{multirow}
\usepackage{pdflscape}
\usepackage{tikz}
\usepackage{comment}
\usepackage{nicefrac}
\usepackage{subcaption}
\usepackage{dsfont}
\usepackage{algorithmic}
\usepackage{algorithm}

\newlength{\figurewidth}
\newlength{\figureheight}
\def\tableskip{\vskip 10pt plus 2pt minus 2pt\relax}
\def\figureskip{\vskip 10pt plus 2pt minus 2pt\relax}

\newtheorem{remark}{Remark}

\def\func#1{\mathop{\rm #1}}

\graphicspath{{Figures/}}

\begin{document}
\title{\textbf{\color{amundi_blue}A Note on Portfolio Optimization\\with Quadratic Transaction Costs}%
\footnote{The authors are grateful to Jules Roche for his helpful comments.}}

\author{{\color{amundi_dark_blue} Pierre Chen} \\
Quantitative Research \\
Amundi Asset Management, Paris \\
\texttt{pierre.chen@amundi.com} \and
{\color{amundi_dark_blue} Edmond Lezmi} \\
Quantitative Research \\
Amundi Asset Management, Paris \\
\texttt{edmond.lezmi@amundi.com} \and
{\color{amundi_dark_blue} Thierry Roncalli} \\
Quantitative Research \\
Amundi Asset Management, Paris \\
\texttt{thierry.roncalli@amundi.com} \and
{\color{amundi_dark_blue} Jiali Xu} \\
Quantitative Research \\
Amundi Asset Management, Paris \\
\texttt{jiali.xu@amundi.com}}
\date{\color{amundi_dark_blue}November 2019}

\maketitle

\begin{abstract}
In this short note, we consider mean-variance optimized portfolios with
transaction costs. We show that introducing quadratic transaction costs makes
the optimization problem more difficult than using linear transaction costs.
The reason lies in the specification of the budget constraint, which is no
longer linear. We provide numerical algorithms for solving this issue and
illustrate how transaction costs may considerably impact the expected returns of
optimized portfolios.
\end{abstract}

\medskip


\section{Introduction}

The general approach for introducing liquidity management in the mean-variance
optimization model of Markowitz (1952) is to assume fixed bid-ask spreads. We
then obtain the linear transaction cost model, which can be solved using an
augmented quadratic programming problem (Scherer, 2007). However, as shown by
Lecesne and Roncoroni (2019a, 2019b), unit transaction costs may be a linear
function of the trading size, implying that a model with quadratic transaction
costs may be more appropriate. In this article, we investigate this approach
and show how linear and quadratic transaction costs modify the mean-variance
optimized framework. In particular, we do not obtain a standard QP problem when
transaction costs are quadratic, because the budget constraint is no longer
linear. In this case, we obtain a quadratically constrained quadratic program
(QCQP), which is an NP-hard problem. However, using the ADMM framework,
we are able to derive an efficient algorithm that solves this issue.
Finally, we use this algorithm to
illustrate the impact of transaction costs on optimized portfolios and
Markowitz efficient frontiers.

\clearpage

\section{Introducing transaction costs into portfolio optimization}

\subsection{Mean-variance optimization with transaction costs}

We consider a universe of $n$ assets. Let $w=\left( w_{1},\ldots
,w_{n}\right) $ be a portfolio. The return of Portfolio $w$ is given by:%
\begin{equation*}
R\left( w\right) =\sum_{i=1}^{n}w_{i}R_{i}=w^{\top }R
\end{equation*}%
where $R=\left( R_{1},\ldots ,R_{n}\right) $ is the random vector of asset
returns. If we note $\mu $ and $\Sigma $ the vector of expected returns and the
covariance matrix of asset returns, we deduce that the expected return
of Portfolio $w$ is equal to:%
\begin{equation*}
\mu \left( w\right) =\mathbb{E}\left[ R\left( w\right) \right] =w^{\top }\mu
\end{equation*}
whereas its variance is given by:%
\begin{equation*}
\sigma ^{2}\left( w\right) =\mathbb{E}\left[ \left( R\left( w\right) -\mu
\left( w\right) \right) ^{2}\right] =w^{\top }\Sigma w
\end{equation*}%
The mean-variance optimization framework of Markowitz (1952) consists in
maximizing the expected return $\mu \left( w\right) $ for a fixed value $%
\sigma ^{\star }$ of the volatility $\sigma \left( w\right) $. This can be
achieved by maximizing the quadratic utility function:%
\begin{equation*}
\mathcal{U}\left( w\right) =\gamma \mu \left( w\right) -\frac{1}{2}\sigma
^{2}\left( w\right)
\end{equation*}%
The mean-variance optimization framework can then be rewritten as a standard
quadratic programming (QP) problem:%
\begin{eqnarray}
w^{\star } &=&\arg \min \frac{1}{2}w^{\top }\Sigma w-\gamma w^{\top }\mu
\label{eq:qp1} \\
&\mathrm{s.t.}&\left\{
\begin{array}{l}
\mathbf{1}_{n}^{\top }w=1 \\
\mathbf{0}_{n}\leq w\leq \mathbf{1}_{n}%
\end{array}%
\right.   \notag
\end{eqnarray}%
The budget constraint $\mathbf{1}_{n}^{\top }w=1$ (or $\sum_{i=1}^{n}w_{i}=1$)
implies that the wealth is entirely invested, whereas the second constraint
indicates that Portfolio $w$ is long-only\footnote{This constraint can be
removed.}.\smallskip

Let us now introduce transaction costs. We note $\tilde{w}$ the current
portfolio and $\mathcal{C}\left( w\mid \tilde{w}\right) $ the cost of
rebalancing the current portfolio $\tilde{w}$ towards the portfolio $w$. We
deduce that the net return is equal to the gross return minus the transaction
costs:%
\begin{equation*}
R\left( w\mid \tilde{w}\right) =R\left( w\right) -\mathcal{C}\left( w\mid
\tilde{w}\right)
\end{equation*}%
It follows that:%
\begin{eqnarray*}
\mu \left( w\mid \tilde{w}\right)  &=&\mathbb{E}\left[ R\left( w\mid \tilde{w%
}\right) \right]  \\
&=&\mu \left( w\right) -\mu _{\mathcal{C}}\left( w\mid \tilde{w}\right)
\end{eqnarray*}%
and:%
\begin{eqnarray*}
\sigma ^{2}\left( w\mid \tilde{w}\right)  &=&\mathbb{E}\left[ \left( R\left(
w\mid \tilde{w}\right) -\mu \left( w\mid \tilde{w}\right) \right) ^{2}\right]
\\
&=&\mathbb{E}\left[ \left( R\left( w\right) -\mu \left( w\right) +\mu _{%
\mathcal{C}}\left( w\mid \tilde{w}\right) -\mathcal{C}\left( w\mid \tilde{w}%
\right) \right) ^{2}\right]  \\
&=&\sigma ^{2}\left( w\right) +\sigma _{\mathcal{C}}^{2}\left( w\mid \tilde{w%
}\right) -2\rho _{\mathcal{C}}\left( w\mid \tilde{w}\right)
\sigma\left( w\right)\sigma _{\mathcal{C}}\left( w\mid \tilde{w%
}\right)
\end{eqnarray*}%
where $\mu _{\mathcal{C}}\left( w\mid \tilde{w}\right) =\mathbb{E}\left[
\mathcal{C}\left( w\mid \tilde{w}\right) \right] $ is the expected cost of
rebalancing and $\sigma _{\mathcal{C}}\left( w\mid \tilde{w}\right) $ is the
standard deviation of $\mathcal{C}\left( w\mid \tilde{w}\right) $. The function
$\rho _{\mathcal{C}}\left( w\mid \tilde{w}\right) $ is the correlation between
the gross return $R\left( w\right) $ and the transaction
cost $\mathcal{C}\left( w\mid \tilde{w}\right) $. Generally, we assume that $%
\rho _{\mathcal{C}}\left( w\mid \tilde{w}\right) \approx 0$. We notice that
transaction costs impact both the expected return and the volatility of the
portfolio. However, this is not the only effect. Indeed, we also have to
finance the rebalancing process since the wealth before and after is not the
same. Therefore, the budget constraint becomes:%
\begin{equation*}
\mathbf{1}_{n}^{\top }w+\mathcal{C}\left( w\mid \tilde{w}\right) =1
\end{equation*}%
Here, we face an issue because the budget constraint is stochastic. This is why
portfolio managers assume that transaction costs are known and not
random. In this case, the optimization problem becomes:%
\begin{eqnarray}
w^{\star } &=&\arg \min \frac{1}{2}w^{\top }\Sigma w-\gamma \left( w^{\top
}\mu -\mathcal{C}\left( w\mid \tilde{w}\right) \right)   \label{eq:qp2} \\
&\mathrm{s.t.}&\left\{
\begin{array}{l}
\mathbf{1}_{n}^{\top }w+\mathcal{C}\left( w\mid \tilde{w}\right) =1 \\
\mathbf{0}_{n}\leq w\leq \mathbf{1}_{n}%
\end{array}%
\right.   \notag
\end{eqnarray}

\subsection{Specification of transaction costs}

\begin{figure}[tbph]
\centering \caption{An example of linear and transaction costs (in \%)}
\label{fig:tc1} \figureskip
\includegraphics[width = \figurewidth, height = \figureheight]{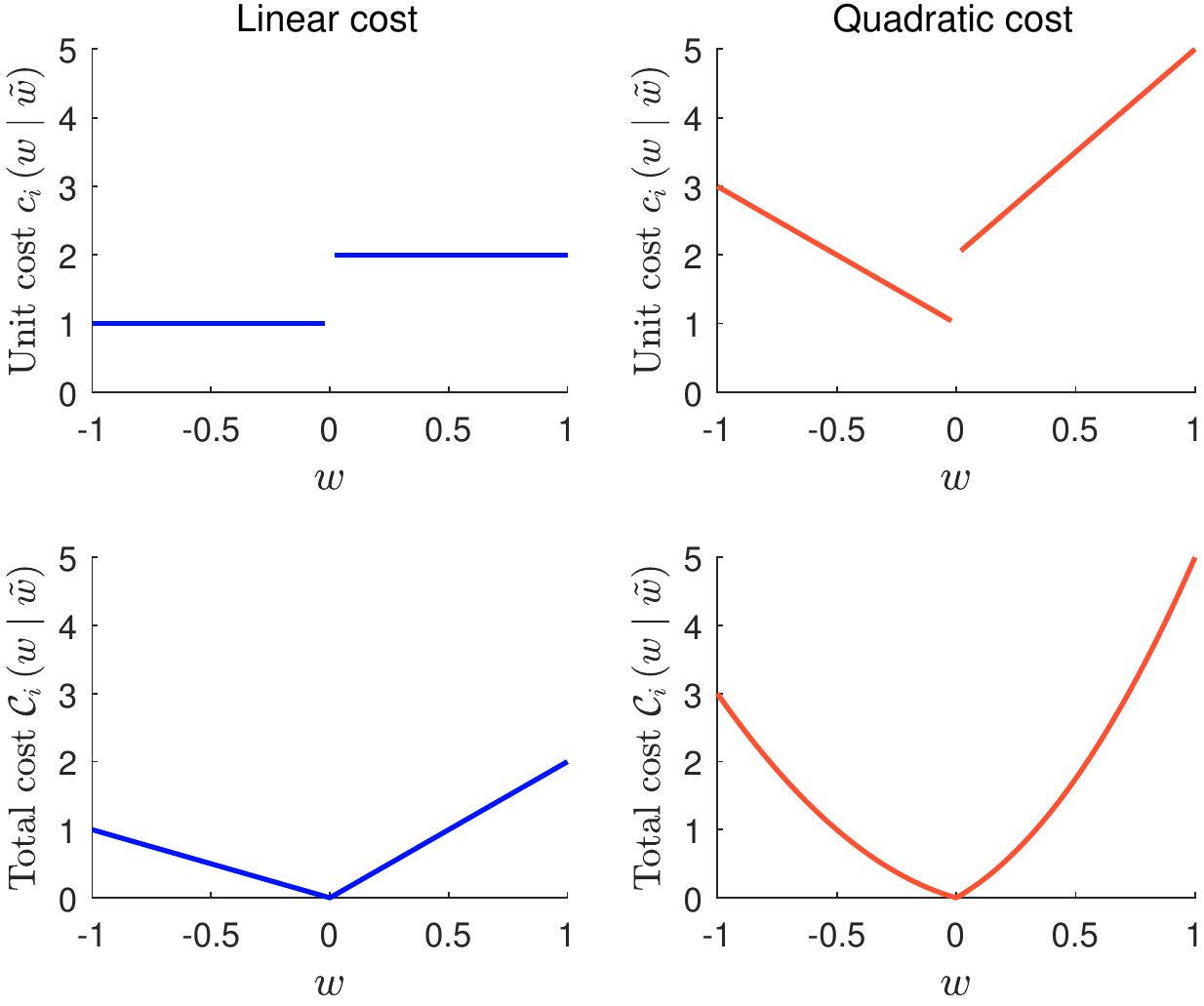}
\end{figure}

A first idea is to consider constant transaction costs. In this case, we
have:%
\begin{equation*}
\mathcal{C}\left( w\mid \tilde{w}\right) =\sum_{i=1}^{n}c_{i}\cdot \left\vert
w_{i}-\tilde{w}_{i}\right\vert
\end{equation*}%
where $c_{i}$ is the unit cost associated with Asset $i$. A better formulation
is to distinguish bid and ask prices. Following Scherer (2007), we have:%
\begin{eqnarray*}
\mathcal{C}\left( w\mid \tilde{w}\right)  &=&\mathcal{C}^{-}\left( w\mid
\tilde{w}\right) +\mathcal{C}^{+}\left( w\mid \tilde{w}\right)  \\
&=&\sum_{i=1}^{n}c_{i}^{-}\cdot \max \left( \tilde{w}_{i}-w_{i},0\right)
+\sum_{i=1}^{n}c_{i}^{+}\cdot \max \left( w_{i}-\tilde{w}_{i},0\right)
\end{eqnarray*}%
where $c_{i}^{-}$ and $c_{i}^{+}$ are the bid and ask unit transaction
costs. We deduce that the transaction cost for Asset $i$ satisfies:%
\begin{equation}
\mathcal{C}_{i}\left( w\mid \tilde{w}\right) =\left\{
\begin{array}{ll}
c_{i}^{-}\cdot \left( \tilde{w}_{i}-w_{i}\right)  & \text{if }w_{i}<\tilde{w}%
_{i} \\
0 & \text{if }w_{i}=\tilde{w}_{i} \\
c_{i}^{+}\cdot \left( w_{i}-\tilde{w}_{i}\right)  & \text{if }w_{i}>\tilde{w}%
_{i}%
\end{array}%
\right.   \label{eq:tc1}
\end{equation}%
In this approach, the unit transaction cost is fixed and does not depend on
the rebalancing weight:%
\begin{equation*}
c_{i}\left( w\mid \tilde{w}\right) =\frac{\mathcal{C}_{i}\left( w\mid
\tilde{w}\right) }{\left\vert w_{i}-\tilde{w}_{i}\right\vert }=\left\{
\begin{array}{ll}
c_{i}^{-} & \text{if }w_{i}<\tilde{w}_{i} \\
0 & \text{if }w_{i}=\tilde{w}_{i} \\
c_{i}^{+} & \text{if }w_{i}>\tilde{w}_{i}%
\end{array}%
\right.
\end{equation*}%
We can also assume that the unit transaction cost is a linear function of
the rebalancing weight:%
\begin{equation*}
c_{i}\left( w\mid \tilde{w}\right) =\left\{
\begin{array}{ll}
c_{i}^{-}+\delta _{i}^{-}\cdot \left( \tilde{w}_{i}-w_{i}\right)  & \text{if
}w_{i}<\tilde{w}_{i} \\
0 & \text{if }w_{i}=\tilde{w}_{i} \\
c_{i}^{+}+\delta _{i}^{+}\cdot \left( w_{i}-\tilde{w}_{i}\right)  & \text{if
}w_{i}>\tilde{w}_{i}%
\end{array}%
\right.
\end{equation*}%
It follows that:%
\begin{equation}
\mathcal{C}_{i}\left( w\mid \tilde{w}\right) =\left\{
\begin{array}{ll}
c_{i}^{-}\cdot \left( \tilde{w}_{i}-w_{i}\right) +\delta _{i}^{-}\cdot
\left( \tilde{w}_{i}-w_{i}\right) ^{2} & \text{if }w_{i}<\tilde{w}_{i} \\
0 & \text{if }w_{i}=\tilde{w}_{i} \\
c_{i}^{+}\cdot \left( w_{i}-\tilde{w}_{i}\right) +\delta _{i}^{+}\cdot
\left( w_{i}-\tilde{w}_{i}\right) ^{2} & \text{if }w_{i}>\tilde{w}_{i}%
\end{array}%
\right.   \label{eq:tc2}
\end{equation}
In the academic literature, Specification (\ref{eq:tc1}) is known under the
term \textquoteleft \textit{linear transaction costs}\textquoteright, whereas
Specification (\ref{eq:tc2}) corresponds to \textquoteleft \textit{quadratic
transaction costs}\textquoteright. An example is provided in Figure
\ref{fig:tc1}, where bid and ask transaction costs are different\footnote{For
instance, in the case of corporate bonds, there are some periods where it is
easier to sell bonds than buy bonds or the contrary.}. On the left side, we
have reported the linear case, whereas the quadratic case corresponds to the
right side\footnote{The parameters are the following: $c_{i}^{-} = 1\%$,
$c_{i}^{+} = 2\%$, $\delta _{i}^{-}= 2\%$ and $\delta _{i}^{+} = 3\%$.
Moreover, we assume that the current allocation $\tilde{w}$ is equal to $0$.}.
We notice that introducing quadratic costs has a more adverse effect on the
portfolio's return. By construction, the choice of one specification will
impact portfolio optimization, especially if the rebalancing is significant.

\section{The case of linear transaction costs}

\subsection{The augmented QP solution}

Since $\mathcal{C}\left( w\mid \tilde{w}\right) $ is a nonlinear function of
$w$, Problem (\ref{eq:qp2}) is not a standard QP problem. This is why Scherer
(2007) suggested rewriting the transaction costs as follows:
\begin{equation*}
\mathcal{C}\left( w\mid \tilde{w}\right) =c_{i}^{-}\cdot \Delta
w_{i}^{-}+c_{i}^{+}\cdot \Delta w_{i}^{+}
\end{equation*}%
where $\Delta w_{i}^{-}=\max \left( \tilde{w}_{i}-w_{i},0\right) $ and $%
\Delta w_{i}^{+}=\max \left( w_{i}-\tilde{w}_{i},0\right) $ represent the sale
and purchase of Asset $i$. By definition, we have $\Delta
w_{i}^{-}\cdot \Delta w_{i}^{+}=0$ and:%
\begin{equation*}
w_{i}=\tilde{w}_{i}+\Delta w_{i}^{+}-\Delta w_{i}^{-}
\end{equation*}%
We deduce that Problem (\ref{eq:qp2}) becomes:
\begin{eqnarray}
w^{\star } &=&\arg \min \frac{1}{2}w^{\top }\Sigma w-\gamma \left(
\sum_{i=1}^{n}w_{i}\cdot \mu _{i}-\sum_{i=1}^{n}\Delta w_{i}^{-}\cdot
c_{i}^{-}-\sum_{i=1}^{n}\Delta w_{i}^{+}\cdot c_{i}^{+}\right)
\label{eq:qp3} \\
&\mathrm{s.t.}&\left\{
\begin{array}{l}
\sum_{i=1}^{n}w_{i}+\sum_{i=1}^{n}\Delta w_{i}^{-}\cdot
c_{i}^{-}+\sum_{i=1}^{n}\Delta w_{i}^{+}\cdot c_{i}^{+}=1 \\
w_{i}+\Delta w_{i}^{-}-\Delta w_{i}^{+}=\tilde{w}_{i} \\
\mathbf{0}_{n}\leq w\leq \mathbf{1}_{n}%
\end{array}%
\right.   \notag
\end{eqnarray}%
We notice that we obtain a QP problem with respect to the variables $x=\left(
w,\Delta w^{-},\Delta w^{+}\right) $. Indeed, we have:
\begin{eqnarray}
x^{\star } &=&\arg \min \frac{1}{2}x^{\top }Qx-x^{\top }R  \label{eq:qp4} \\
&\mathrm{s.t.}&\left\{
\begin{array}{l}
Ax=B \\
x^{-}\leq x\leq x^{+}%
\end{array}%
\right.   \notag
\end{eqnarray}%
where:%
\begin{equation*}
Q=\left(
\begin{array}{ccc}
\Sigma  & \mathbf{0}_{n,n} & \mathbf{0}_{n,n} \\
\mathbf{0}_{n,n} & \mathbf{0}_{n,n} & \mathbf{0}_{n,n} \\
\mathbf{0}_{n,n} & \mathbf{0}_{n,n} & \mathbf{0}_{n,n}%
\end{array}%
\right)
\end{equation*}%
and:%
\begin{equation*}
R=\gamma \left(
\begin{array}{c}
\mu  \\
-c^{-} \\
-c^{+}%
\end{array}%
\right)
\end{equation*}%
For the equality constraint, we obtain:
\begin{equation*}
A=\left(
\begin{array}{ccc}
\mathbf{1}_{n}^{\top } & \left( c^{-}\right) ^{\top } & \left( c^{+}\right)
^{\top } \\
I_{n} & I_{n} & -I_{n}%
\end{array}%
\right)
\end{equation*}%
and:%
\begin{equation*}
B=\left(
\begin{array}{c}
1 \\
\tilde{w}%
\end{array}%
\right)
\end{equation*}%
For the bounds, we notice that:
\begin{eqnarray*}
0\leq w_{i}\leq 1 &\Leftrightarrow &0\leq \tilde{w}_{i}+\Delta
w_{i}^{+}-\Delta w_{i}^{-}\leq 1 \\
&\Leftrightarrow &-\tilde{w}_{i}\leq \Delta w_{i}^{+}-\Delta w_{i}^{-}\leq 1-%
\tilde{w}_{i} \\
&\Leftrightarrow &\left\{
\begin{array}{ll}
-\tilde{w}_{i}\leq \Delta w_{i}^{+}\leq 1-\tilde{w}_{i} & \text{if }\Delta
w_{i}^{-}=0 \\
\tilde{w}_{i}-1\leq \Delta w_{i}^{-}\leq \tilde{w}_{i} & \text{if }\Delta
w_{i}^{+}=0%
\end{array}%
\right.
\end{eqnarray*}%
However, we know that $\Delta w_{i}^{-}\geq 0$ and $\Delta w_{i}^{+}\geq 0$. We
deduce that $x^{-}=\mathbf{0}_{3n}$ and:
\begin{equation*}
x^{+}=\left(
\begin{array}{c}
\mathbf{1}_{n} \\
\tilde{w} \\
\mathbf{1}_{n}-\tilde{w}%
\end{array}%
\right)
\end{equation*}%
Problem (\ref{eq:qp4}) is called an augmented QP problem (Roncalli, 2013),
because we have augmented the number of variables in order to find the optimal
solution $w^{\star }$ which is given by the following relationship:
\begin{equation*}
w^{\star }=\left(
\begin{array}{ccc}
I_{n} & \mathbf{0}_{n,n} & \mathbf{0}_{n,n}%
\end{array}%
\right) x^{\star }
\end{equation*}

\subsection{The efficient frontier with linear transaction costs}

We consider an investment universe of $7$ assets. Their expected return and
volatility expressed as a $\%$ are equal to:
\begin{equation*}
\begin{tabular}{c|ccccccc}
\hline
$i$ & 1 & 2 & 3 & 4 & 5 & 6 & 7 \\ \hline
$\mu_i$    & 1.00 & 2.00 & 3.00 & 4.00 & 5.00 & 7.50 & 10.00 \\
$\sigma_i$ & 1.00 & 2.00 & 3.00 & 4.00 & 5.00 & 7.50 & 10.00 \\
\hline
\end{tabular}
\end{equation*}
We also consider a constant correlation matrix of $25\%$ between asset returns.
The initial portfolio is composed of $50\%$ of Asset 1 and $50\%$ of Asset
2.\smallskip

\begin{figure}[tbph]
\centering
\caption{Efficient frontier $\left( \sigma \left(
\bar{w}^{\star }\right) ,\mu _{\mathrm{net}}\left( \bar{w}^{\star }\right)
\right) $ with $c^{-} = 20$ bps and $c^{+} = 10$ bps}
\label{fig:lc_mvo1}
\figureskip
\includegraphics[width = \figurewidth, height = \figureheight]{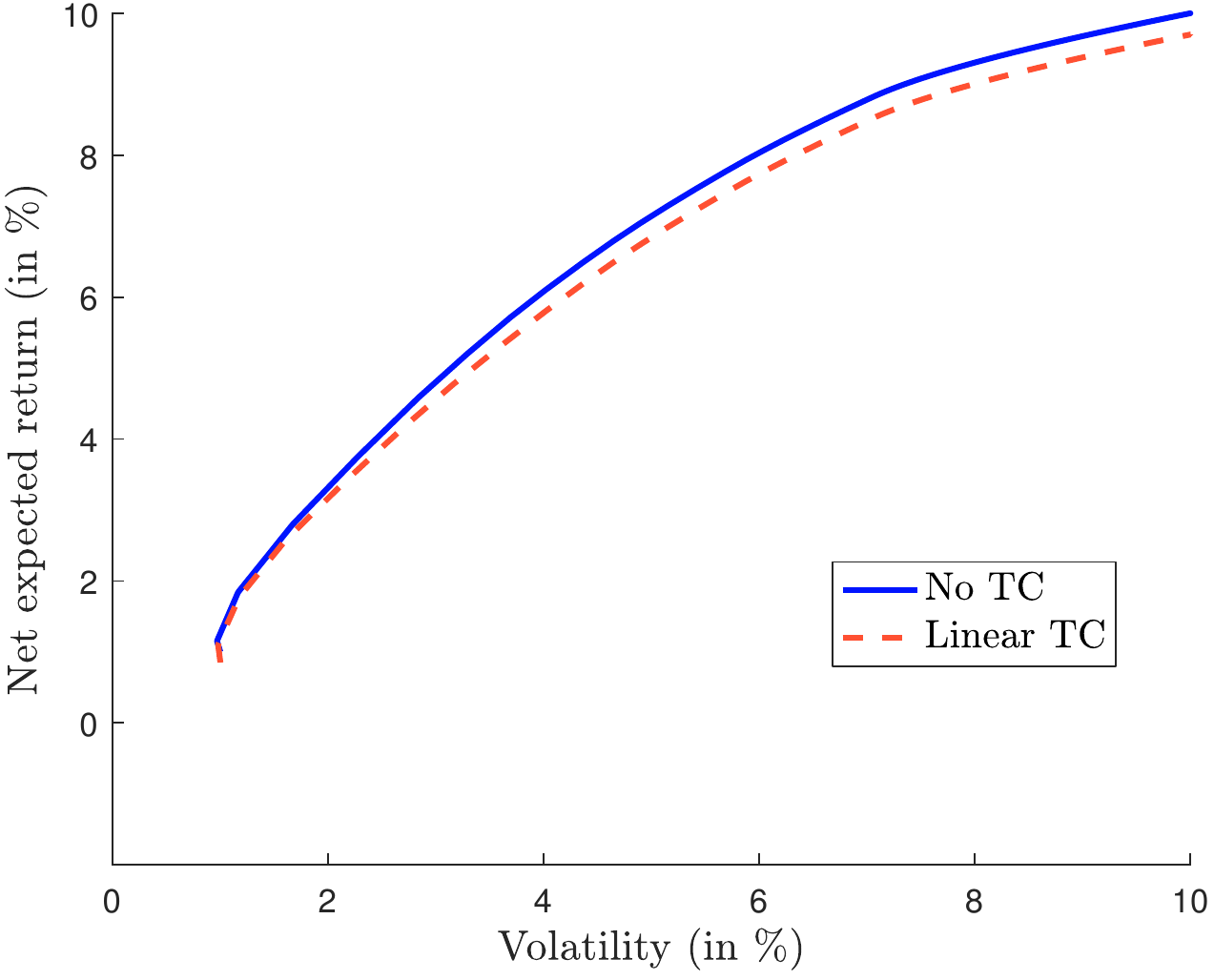}
\end{figure}

By assuming fixed transaction costs $c^{-}=20$ bps and $c^{+}=10$ bps, we
obtain the efficient frontier that is reported in Figure \ref{fig:lc_mvo1}.
Here, we face an issue because transaction costs imply that
$\sum_{i=1}^{n}w_{i}^{\star }<1$. Therefore, the efficient frontier cannot be
represented by the pair $\left( \sigma \left( w^{\star }\right) ,\mu \left(
w^{\star }\right) \right) $ because the net wealth $\sum_{i=1}^{n}w_{i}^{\star
}$ depends on the values taken by $c^{-}$ and $c^{+}$. With no transaction
costs, we retrieve the classical efficient frontier of Markowitz (1952).
However, in order to compare efficient frontiers, we have to normalize the
optimized portfolio:
\begin{equation*}
\bar{w}^{\star }=\frac{w_{i}^{\star }}{\sum_{i=1}^{n}w_{i}^{\star }}
\end{equation*}%
Indeed, plotting $\left( \sigma \left( w^{\star }\right) ,\mu \left( w^{\star
}\right) \right) $ is misleading since we have paid transaction costs in order
to rebalance the portfolio. For instance, if the transaction costs are high, we
have $\sum_{i=1}^{n}w_{i}^{\star }\ll 1$ and we may obtain a very low
volatility and some optimized portfolios may be on the left of the Markowitz
efficient frontier. The reason is that the portfolio is less risky on a nominal
basis because the portfolio notional is reduced. This is why it is better to
consider the expected return adjusted by the transaction costs (also called the
`\textit{net expected return}'), which is equal to $\mu _{\mathrm{net}}\left(
\bar{w}^{\star }\right) =\mu \left( \bar{w}^{\star }\right) -
\mathcal{C}\left(w^{\star} \mid \tilde{w}\right)$. The efficient frontier with
transaction costs is then represented by the curve $\left( \sigma \left(
\bar{w}^{\star }\right) ,\mu _{\mathrm{net}}\left( \bar{w}^{\star }\right)
\right) $. However, with $c^{-}=20$ bps and $c^{+}=10$ bps, Figure
\ref{fig:lc_mvo1} gives the impression that transaction costs have little
impact on the efficient frontier.\smallskip

\begin{figure}[tbph]
\centering
\caption{Efficient frontier $\left( \sigma \left(
\bar{w}^{\star }\right) ,\mu _{\mathrm{net}}\left( \bar{w}^{\star }\right)
\right) $ with $c^{-} = 2\%$ and $c^{+} = 1\%$}
\label{fig:lc_mvo2}
\figureskip
\includegraphics[width = \figurewidth, height = \figureheight]{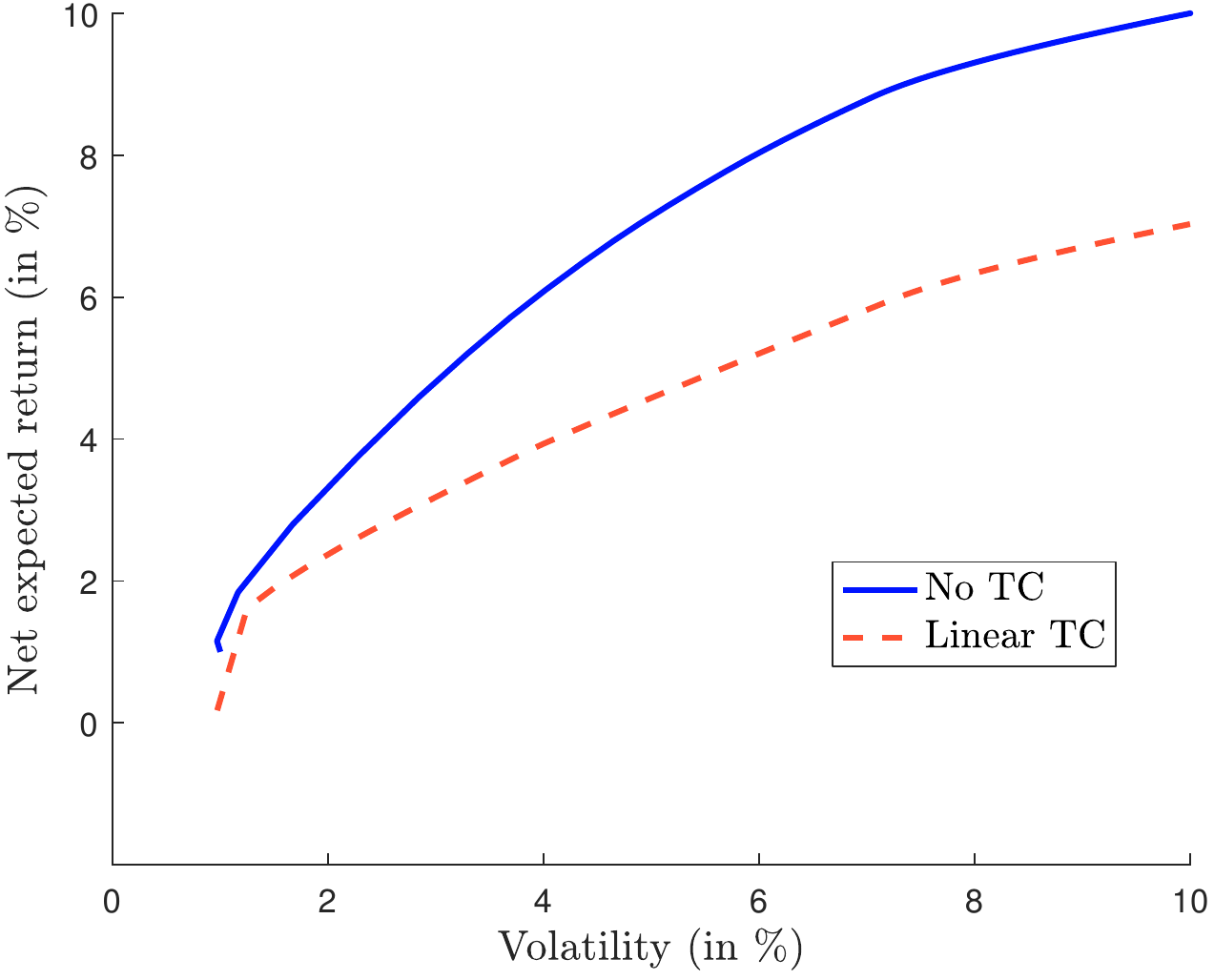}
\end{figure}

\begin{figure}[tbph]
\centering
\caption{Transaction cost}
\label{fig:lc_mvo3}
\figureskip
\includegraphics[width = \figurewidth, height = \figureheight]{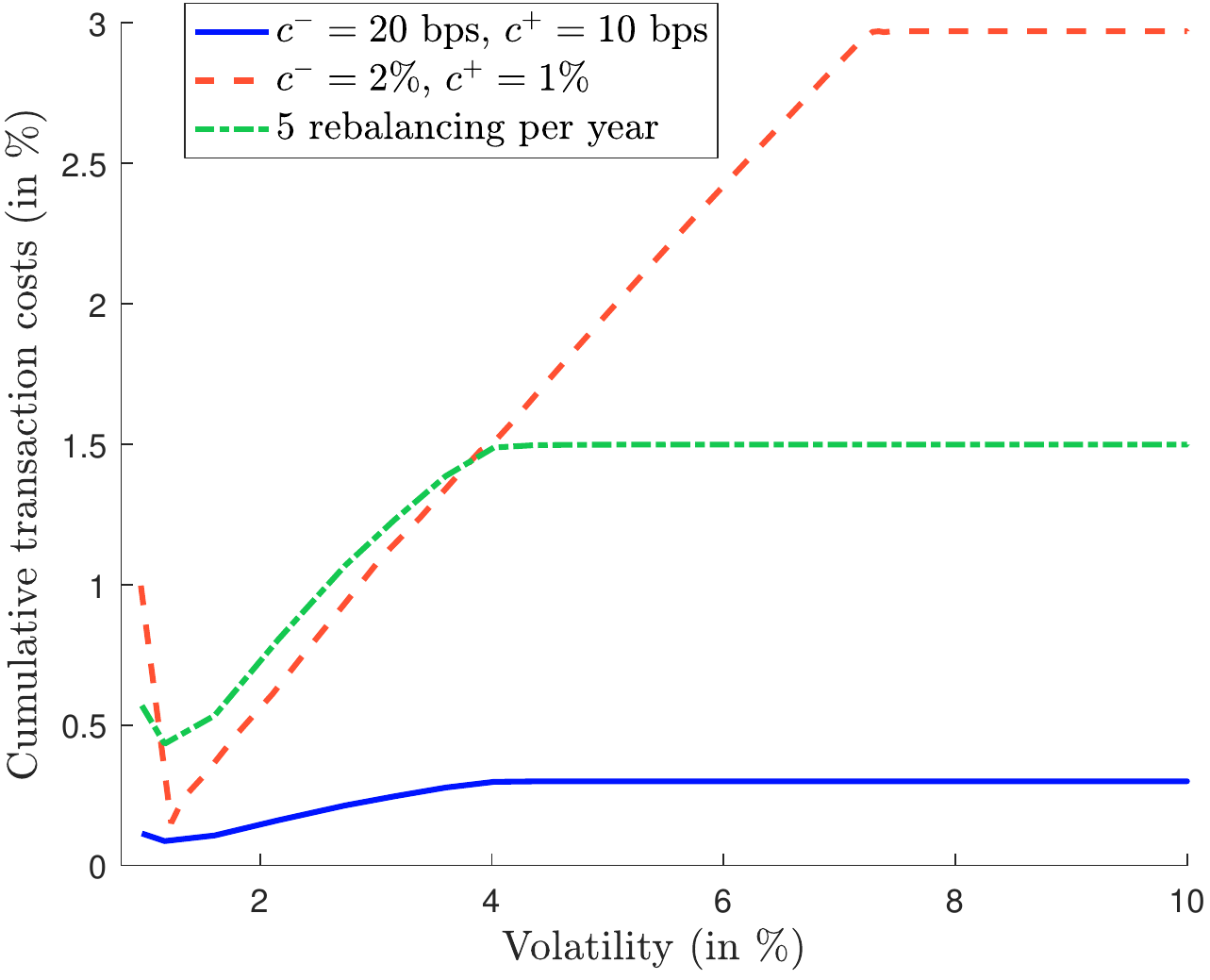}
\end{figure}

Let us now consider an unrealistic case: $c^{-} = 2\%$ and $c^{+} = 1\%$. We
obtain Figure \ref{fig:lc_mvo2} and we notice the big impact of transaction
costs on the expected return of the portfolio. In Figure \ref{fig:lc_mvo3}, we
have reported the total amount of transaction costs with respect to portfolio
volatility. Since the case $c^{-} = 20$ bps and $c^{+} = 10$ bps is more
realistic, it may not reflect the real impact on a trading strategy. Indeed,
these transaction costs are paid at each rebalancing date. The efficient
frontier considers a yearly expected return, whereas the net expected return
assumes only one portfolio rebalancing in the year, and does not take into
account the total turnover of the portfolio. For example, if we assume that we
rebalance the portfolio $5$ times in the year, we obtain the green curve that
illustrates how cumulative transaction costs can be damaging for portfolio
performance.

\section{Introducing quadratic transaction costs}

\subsection{The issue of the quadratic budget constraint}

In the case of quadratic transaction costs, we can use the same approach by
considering augmented variables. We deduce that:
\begin{eqnarray*}
\mathcal{C}\left( w\mid \tilde{w}\right)  &=&\sum_{i=1}^{n}\Delta
w_{i}^{-}\left( c_{i}^{-}+\delta _{i}^{-}\Delta w_{i}^{-}\right)
+\sum_{i=1}^{n}\Delta w_{i}^{+}\left( c_{i}^{+}+\delta _{i}^{+}\Delta
w_{i}^{+}\right)  \\
&=&\Delta w^{-\top }c^{-}+\Delta w^{-\top }\Delta ^{-}\Delta w^{-}+\Delta
w^{+\top }c^{+}+\Delta w^{+\top }\Delta ^{+}\Delta w^{+}
\end{eqnarray*}%
where $\Delta ^{-}=\func{diag}\left( \delta _{1}^{-},\ldots ,\delta
_{n}^{-}\right) $ and $\Delta ^{+}=\func{diag}\left( \delta _{1}^{+},\ldots
,\delta _{n}^{+}\right) $ are two diagonal matrices. It follows that the
objective function of Problem (\ref{eq:qp2}) remains quadratic:%
\begin{eqnarray*}
f\left( w,\Delta w^{-},\Delta w^{+}\right)  &=&\frac{1}{2}w^{\top }\Sigma
w-\gamma \left( w^{\top }\mu -\mathcal{C}\left( w\mid \tilde{w}\right)
\right)  \\
&=&\frac{1}{2}\left( w^{\top }\Sigma w+\Delta w^{-\top }\left( 2\gamma \Delta
^{-}\right) \Delta w^{-}+\Delta w^{+\top }\left( 2\gamma \Delta
^{+}\right) \Delta w^{+}\right) - \\
&&\gamma \left( w^{\top }\mu -\Delta w^{-\top }c^{-}-\Delta w^{+\top
}c^{+}\right)
\end{eqnarray*}%
but the budget constraint is no longer linear:%
\begin{equation*}
\underset{\text{Linear term}}{\underbrace{\mathbf{1}_{n}^{\top }w+\Delta
w^{-\top }c^{-}+\Delta w^{+\top }c^{+}}}+\underset{\text{Quadratic term}}{%
\underbrace{\Delta w^{-\top }\Delta ^{-}\Delta w^{-}+\Delta w^{+\top }\Delta
^{+}\Delta w^{+}}}=1
\end{equation*}%
Indeed, the budget constraint is composed of a linear term and a quadratic
term.\smallskip

Let $x=\left( w,\Delta w^{-},\Delta w^{+}\right) $ be the vector of original
variables and augmented variables. We obtain:
\begin{eqnarray}
x^{\star } &=&\arg \min \frac{1}{2}x^{\top }Qx-x^{\top }R \label{eq:qp5}\\
&\mathrm{s.t.}&\left\{
\begin{array}{l}
A_{1}x+x^{\top }C_1x=B_{1} \\
A_{2}x=B_{2} \\
x^{-}\leq x\leq x^{+}%
\end{array}%
\right.   \notag
\end{eqnarray}%
where:%
\begin{equation*}
Q=\left(
\begin{array}{ccc}
\Sigma  & \mathbf{0}_{n,n} & \mathbf{0}_{n,n} \\
\mathbf{0}_{n,n} & 2\gamma \Delta ^{-} & \mathbf{0}_{n,n} \\
\mathbf{0}_{n,n} & \mathbf{0}_{n,n} & 2\gamma \Delta ^{+}%
\end{array}%
\right)
\end{equation*}%
and:%
\begin{equation*}
R=\gamma \left(
\begin{array}{c}
\mu  \\
-c^{-} \\
-c^{+}%
\end{array}%
\right)
\end{equation*}%
For the equality constraints, we obtain:
\begin{equation*}
\left(
\begin{array}{c}
A_{1} \\
A_{2}%
\end{array}%
\right) =\left(
\begin{array}{ccc}
\mathbf{1}_{n}^{\top } & \left( c^{-}\right) ^{\top } & \left( c^{+}\right)
^{\top } \\
I_{n} & I_{n} & -I_{n}%
\end{array}%
\right)
\end{equation*}%
and:%
\begin{equation*}
\left(
\begin{array}{c}
B_{1} \\
B_{2}%
\end{array}%
\right) =\left(
\begin{array}{c}
1 \\
\tilde{w}%
\end{array}%
\right)
\end{equation*}%
The matrix $C_1$ is defined as follows:%
\begin{equation*}
C_1=\left(
\begin{array}{ccc}
\mathbf{0}_{n,n} & \mathbf{0}_{n,n} & \mathbf{0}_{n,n} \\
\mathbf{0}_{n,n} & \Delta ^{-} & \mathbf{0}_{n,n} \\
\mathbf{0}_{n,n} & \mathbf{0}_{n,n} & \Delta ^{+}%
\end{array}%
\right)
\end{equation*}
The bounds remain the same. We have $x^{-}=\mathbf{0}_{3n}$ and:
\begin{equation*}
x^{+}=\left(
\begin{array}{c}
\mathbf{1}_{n} \\
\tilde{w} \\
\mathbf{1}_{n}-\tilde{w}%
\end{array}%
\right)
\end{equation*}%
Again, the optimal solution $w^{\star }$ is given by the following
relationship:
\begin{equation*}
w^{\star }=\left(
\begin{array}{ccc}
I_{n} & \mathbf{0}_{n,n} & \mathbf{0}_{n,n}%
\end{array}%
\right) x^{\star }
\end{equation*}

\subsection{The ADMM solution}

Following Perrin and Roncalli (2019), we can use the alternating direction
method of multipliers (ADMM) algorithm formulated by Gabay and Mercier (1976)
to solve Problem (\ref{eq:qp5}) and overcome the non linear constraint. To do
this, we leave the objective function as well as all the linear constraints in
the $x$-update and put the non linear constraint in the $y$-update. In this
case, the $x$-update is easily solved using QP, but the $y$-update is an
NP-hard problem in the general case.

\subsubsection{The ADMM formulation}

Problem (\ref{eq:qp5}) is equivalent to:
\begin{eqnarray}
\left\{ x^{\star },y^{\star }\right\}  &=&\arg \min_{\left( x,y\right)
}f_{x}\left( x\right) +f_{y}\left( y\right)   \label{eq:admm1} \\
&\text{s.t.}&x-y=\mathbf{0}_{n}  \notag
\end{eqnarray}%
where:%
\begin{equation*}
f_{x}\left( x\right) =\frac{1}{2}x^{\top }Qx-x^{\top }R+\mathds{1}_{\Omega
_{x}}(x)
\end{equation*}
and:%
\begin{equation*}
f_{y}\left( y\right) =\mathds{1}_{\Omega _{y}}(y)
\end{equation*}%
The sets $\Omega _{x}$ and $\Omega _{y}$ are defined as follows:%
\begin{equation*}
\Omega _{x}\left( x\right) =\left\{ x\in \left[ 0,1\right]
^{n}:A_{2}x=B_{2},x^{-}\leq x\leq x^{+}\right\}
\end{equation*}%
and:%
\begin{equation*}
\Omega _{y}\left( y\right) =\left\{ y\in \left[ 0,1\right] ^{n}:A_{1}y+y^{\top
}C_1y=B_{1}\right\}
\end{equation*}
The corresponding ADMM algorithm consists of the following three steps (Boyd
\textsl{et al.}, 2011; Perrin and Roncalli, 2019):

\begin{enumerate}
\item The $x$-update is:
\begin{equation}
x^{\left( k+1\right) }=\arg \min_{x}\left\{ f_{x}\left( x\right) +\frac{%
\varphi }{2}\left\Vert x-y^{\left( k\right) }+u^{\left( k\right) }\right\Vert
_{2}^{2}\right\}   \label{eq:admm2a}
\end{equation}

\item The $y$-update is:%
\begin{equation}
y^{\left( k+1\right) }=\arg \min_{y}\left\{ f_{y}\left( y\right) +\frac{%
\varphi }{2}\left\Vert x^{\left( k+1\right) }-y+u^{\left( k\right) }\right\Vert
_{2}^{2}\right\}   \label{eq:admm2b}
\end{equation}

\item The $u$-update is:%
\begin{equation}
u^{\left( k+1\right) }=u^{\left( k\right) }+x^{\left( k+1\right) }-y^{\left(
k+1\right) }  \label{eq:admm2c}
\end{equation}
\end{enumerate}

\noindent As noted by Perrin and Roncalli (2019), the $x$-update is a QP
problem:%
\begin{eqnarray}
x^{\left( k+1\right) } &=&\arg \min \frac{1}{2}x^{\top }\left( Q+\varphi
I_{3n}\right) x-x^{\top }\left( R+\varphi \left( y^{\left( k\right)
}-u^{\left( k\right) }\right) \right)  \\
&\mathrm{s.t.}&\left\{
\begin{array}{l}
A_{2}x=B_{2} \\
x^{-}\leq x\leq x^{+}%
\end{array}%
\right.   \notag
\end{eqnarray}%
There is no difficulty in finding the numerical solution $x^{\left( k+1\right)
} $. In fact, the issue concerns the calculation of $y^{\left( k+1\right) }$.

\subsubsection{The case $\protect\delta _{i}^{-}=\protect\delta ^{-}$ and $%
\protect\delta _{i}^{+}=\protect\delta ^{+}$}

Generally, the $y$-update is easily solved by combining proximal operators and
the Dykstra algorithm. However, in our case, we cannot use such a decomposition
because the constraint is unusual. In fact, we have the following optimization
problem:
\begin{eqnarray*}
y^{(k+1)} &=&\arg \min_{y}\frac{1}{2}\left\Vert y-v_{y}^{\left( k+1\right)
}\right\Vert _{2}^{2} \\
\text{} &\text{s.t.}&y\in \Omega _{y}
\end{eqnarray*}%
where $v_{y}^{\left( k+1\right) }=x^{\left( k+1\right) }+u^{\left( k\right) }
$. We deduce that the Lagrange function is equal to:
\begin{eqnarray*}
\mathcal{L}\left( y,\lambda \right)  &=&\frac{1}{2}\left\Vert
y-v_{y}^{\left( k+1\right) }\right\Vert _{2}^{2}+ \\
&&\lambda \left( \sum_{i=1}^{n}\left( w_{i}+\Delta w_{i}^{-}\left(
c_{i}^{-}+\delta _{i}^{-}\Delta w_{i}^{-}\right) +\Delta w_{i}^{+}\left(
c_{i}^{+}+\delta _{i}^{+}\Delta w_{i}^{+}\right) \right) -1\right)
\end{eqnarray*}%
Using the similar partition $v_{y}^{\left( k+1\right) }=\left( v,\Delta
v^{-},\Delta v^{+}\right) $ as $y=\left( w,\Delta w^{-},\Delta w^{+}\right) $%
, the KKT conditions are:%
\begin{equation*}
\left\{
\begin{array}{l}
w_{i}-v_{i}+\lambda =0 \\
\Delta w_{i}^{-}-\Delta v_{i}^{-}+\lambda \left( c_{i}^{-}+2\delta
_{i}^{-}\Delta w_{i}^{-}\right) =0 \\
\Delta w_{i}^{+}-\Delta v_{i}^{+}+\lambda \left( c_{i}^{+}+2\delta_i
^{+}y_{i}^{+}\right) =0 \\
\sum_{i=1}^{n}\left( w_{i}+\Delta w_{i}^{-}\left( c_{i}^{-}+\delta
_{i}^{-}\Delta w_{i}^{-}\right) +\Delta w_{i}^{+}\left( c_{i}^{+}+\delta
_{i}^{+}\Delta w_{i}^{+}\right) \right) =1%
\end{array}%
\right.
\end{equation*}%
We then get a nonlinear system of $3n+1$ equations. We first consider the case
$\delta _{i}^{-}=\delta ^{-}$ and $\delta _{i}^{+}=\delta ^{+}$. In Appendix
\ref{appendix:y-update1} on page \pageref{appendix:y-update1},
we show that $\lambda $ is the solution of a quintic equation:%
\begin{equation*}
\alpha _{5}\lambda ^{5}+\alpha _{4}\lambda ^{4}+\alpha _{3}\lambda
^{3}+\alpha _{2}\lambda ^{2}+\alpha _{1}\lambda +\alpha _{0}=0
\end{equation*}%
From this, we can conclude that there are as many solutions to the nonlinear
system as there are real roots to the last polynomial equation. Since we know
that KKT conditions are necessary, it is sufficient to compare the different
solutions obtained for this system in order to find the solution of our
original program\footnote{It is also possible that there are cases where we can
get several solutions as we are projecting onto a quadratic equation. For
example, we would get an infinite number of solutions if we project a point
onto a circle, where this point is its center. However, in the general case, we
avoid these critical points and find only one single real root to the
polynomial equation.}. More general methods are available in order to
numerically solve the nonlinear system such as the Newton-Raphson algorithm.
However, for these methods, it is usually necessary to compute the inverse of a
Hessian matrix at each step of iteration which is very costly (around
$\mathcal{O}\left(
(3n+1)^{3}\right) $). By taking advantage of the derivation of the $y$%
-update, we only need one step of cost $\mathcal{O}\left( 5^{3}\right) $ to
compute the roots of the polynomial in order to solve the system.

\subsubsection{The case $\protect\delta _{i}^{-}\neq \protect\delta _{j}^{-}$
and $\protect\delta _{i}^{+}\neq \protect\delta _{j}^{+}$}

The case $\delta _{i}^{-}\neq \delta _{j}^{-}$ and $\delta _{i}^{+}\neq \delta
_{j}^{+}$ complicates the problem. Indeed, we obtain a
polynomial equation of degree $2n+1$. Another solution is to rewrite the $%
y$-update problem in a matrix form\footnote{$\varphi $ is set to one because
its value has no impact on the solution.}:%
\begin{eqnarray*}
y^{\left( k+1\right) } &=&\arg \min \frac{1}{2}\left( y-v_{y}^{\left(
k+1\right) }\right) ^{\top }\left( y-v_{y}^{\left( k+1\right) }\right)  \\
&\mathrm{s.t.}&A_{1}y+y^{\top }C_1y-B_{1}=0
\end{eqnarray*}%
We obtain a quadratically constrained quadratic program (QCQP). Since a
quadratic equality is not convex, the optimization problem is not convex. More
generally, a QCQP is an NP-hard problem. A numerical solution is therefore to
consider an interior-point algorithm by specifying the gradient of the
objective function, the gradient of the equality constraint and the
Hessian of the Lagrangian\footnote{%
They are respectively equal to $y-v_{y}^{\left( k+1\right) }$, $A_{1}+2C_1 y$ and
$I_{n}+2\lambda C_1$.}. However, since we have only one constraint and the objective function
is simple, we can derive the numerical solution (Park and Boyd, 2017), which is described
in Appendix \ref{appendix:y-update2} on page \pageref{appendix:y-update2}.

\begin{remark}
The ADMM formulation has allowed us to split the QCQP Problem (\ref{eq:qp5}) with two inequality
and two equality constraints into a QP problem ($x$-update)
and a QCQP problem with only one constraint ($y$-update). As explained by Park and Boyd (2017),
solving QCQP with one constraint is feasible and relatively easy.
This is not always the case when there are two or more constraints.
\end{remark}

\subsection{The efficient frontier with quadratic transaction costs}

We consider our previous example. We assume that the current portfolio is the
optimal portfolio $\tilde{w}$ corresponding to volatility of $2\%$. In a second
period, the portfolio manager would increase portfolio risk and target
volatility equal to $4\%$. Portfolio $w_{\mathrm{MVO}}^{\star }$ is the optimal
solution if we do not take into account transaction costs. However, this
portfolio is not realistic if we consider transaction costs. We set
$c^{-}=2\%$, $c^{+}=1\%$, $\delta ^{-}=5\%$ and $\delta ^{+}=5\%$. The results
are given in Table \ref{tab:qc_mvo1}. In the case of linear transaction costs,
we obtain Portfolio $w_{\mathrm{LC}}^{\star }$. We observe that the two
solutions $w_{\mathrm{MVO}}^{\star }$ and $w_{\mathrm{LC}}^{\star }$ are very
different. For instance, the LC solution keeps a significant proportion of
Asset 2 in order to pay less transaction costs. Indeed, Portfolios
$w_{\mathrm{MVO}}^{\star }$ and $w_{\mathrm{LC}}^{\star }$ pay respectively
$1.58\%$ and $0.98\%$ of transaction costs. In the case of quadratic
transaction costs, the solution is Portfolio $w_{\mathrm{QC}}^{\star }$. We
notice that it has a lower turnover than the two previous portfolios. Moreover,
it selects assets with a high return in order to compensate for the transaction
costs. This is why we obtain a weight of $29.13\%$ for Asset 7.\smallskip

\begin{table}[t]
\centering
\caption{Comparison of optimized portfolios with linear and quadratic costs}
\label{tab:qc_mvo1}
\tableskip
\begin{tabular}{c|rrrrrr}
\hline
Asset & \multicolumn{1}{c}{$\tilde{w}$} & \multicolumn{1}{c}{$w_{\mathrm{MVO}}^{\star }$} &
\multicolumn{1}{c}{$w_{\mathrm{LC}}^{\star }$} & \multicolumn{1}{c}{$w_{\mathrm{QC}}^{\star }$} &
\multicolumn{1}{c}{$\bar{w}_{\mathrm{LC}}^{\star }$} & \multicolumn{1}{c}{$\bar{w}_{\mathrm{QC}}^{\star }$} \\
\hline
$1$    & $26.16$ & $ 0.01$ & $ 0.00$ & $ 6.70$ & $ 0.00$ & $ 6.80$ \\
$2$    & $21.41$ & $ 0.08$ & $14.52$ & $10.84$ & $14.67$ & $11.01$ \\
$3$    & $16.13$ & $10.92$ & $16.13$ & $14.32$ & $16.28$ & $14.53$ \\
$4$    & $12.79$ & $22.42$ & $12.79$ & $12.78$ & $12.91$ & $12.98$ \\
$5$    & $10.56$ & $24.77$ & $10.56$ & $10.56$ & $10.67$ & $10.72$ \\
$6$    & $ 7.34$ & $22.59$ & $18.27$ & $14.17$ & $18.45$ & $14.38$ \\
$7$    & $ 5.62$ & $19.22$ & $26.74$ & $29.13$ & $27.01$ & $29.57$ \\ \hline
$\mu \left( w\right)$                                    &  $3.33$ &  $6.08$ &  $5.86$ &  $5.73$ &  $5.92$ &  $5.82$ \\
$\sigma \left( w\right)$                                 &  $2.00$ &  $4.00$ &  $4.00$ &  $4.00$ &  $4.04$ &  $4.06$ \\
$\mathcal{C}_{\mathrm{LC}}\left( w\mid \tilde{w}\right)$ &         &  $1.58$ &  $0.98$ &  $0.94$ &  $    $ &  $    $ \\
$\mathcal{C}_{\mathrm{QC}}\left( w\mid \tilde{w}\right)$ &         &  $2.52$ &  $1.63$ &  $1.49$ &  $    $ &  $    $ \\
$\mu_{\mathrm{LC}}\left( w\mid \tilde{w}\right)$         &  $3.33$ &  $4.50$ &  $4.88$ &  $4.79$ &  $    $ &  $    $ \\
$\mu_{\mathrm{QC}}\left( w\mid \tilde{w}\right)$         &  $3.33$ &  $3.56$ &  $4.23$ &  $4.24$ &  $    $ &  $    $ \\ \hline
\end{tabular}
\end{table}

\begin{figure}[!h]
\centering
\caption{Efficient frontier $\left( \sigma \left(
\bar{w}^{\star }\right) ,\mu _{\mathrm{net}}\left( \bar{w}^{\star }\right)
\right) $ with quadratic transaction costs}
\label{fig:qc_mvo4}
\figureskip
\includegraphics[width = \figurewidth, height = \figureheight]{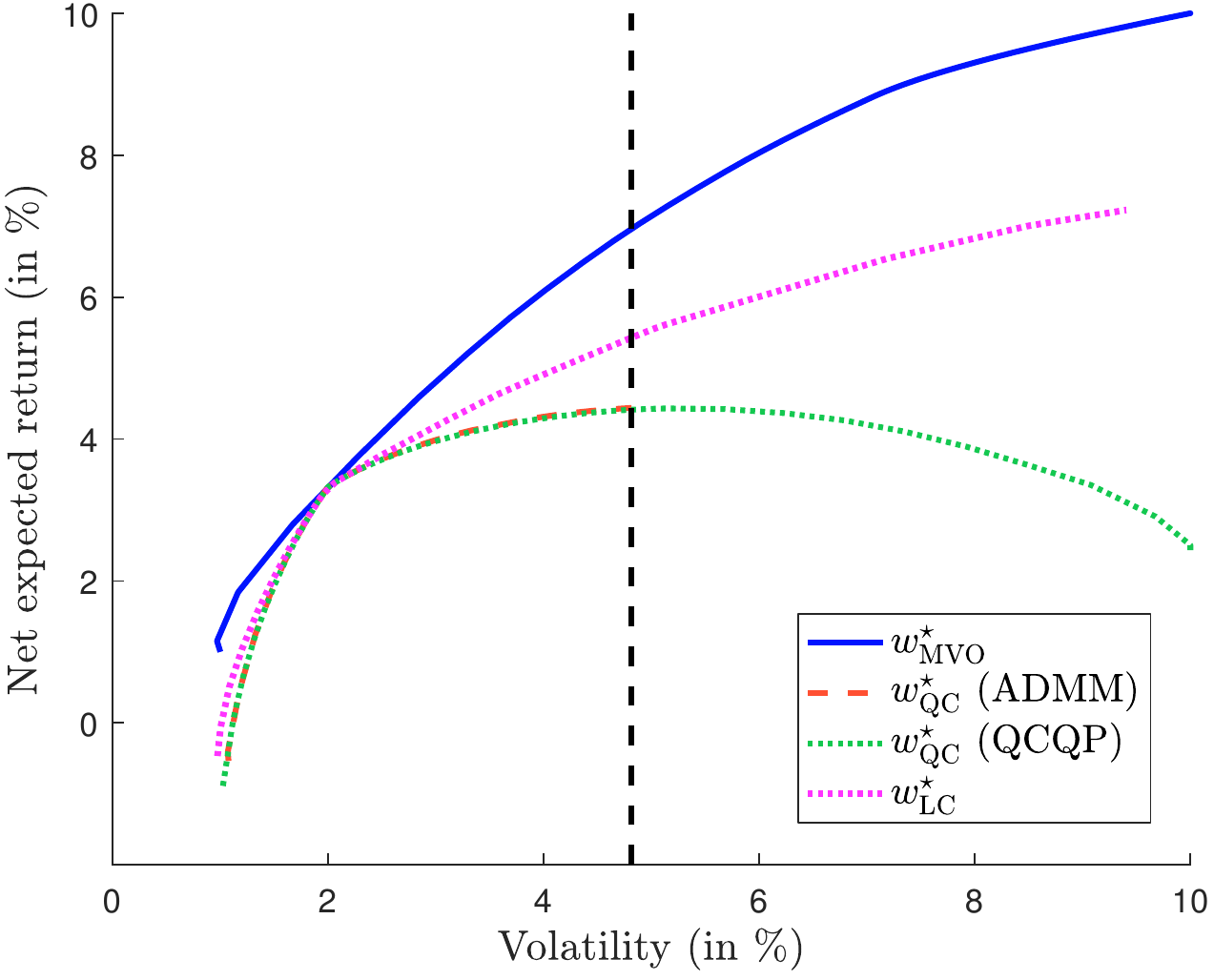}
\end{figure}

In Figure \ref{fig:qc_mvo4}, we have reported the efficient frontier with the
previous transaction costs. We verify that it is below the unconstrained MVO
efficient frontier. We also notice that quadratic transaction costs have a
significant adverse effect on the net expected return of optimized portfolios.
In our example, we rebalance a portfolio, which has low volatility. This
implies that it is inefficient to target an optimized portfolio with high
volatility, because we have to pay substantial transaction costs. For instance,
it is impossible to target an expected return greater than $4\%$, implying that
having portfolio volatility higher than $5\%$ is not optimal.

\begin{remark}
We do not say that it is impossible to target volatility higher than $5\%$, but
we say that it is not optimal. Indeed, the ADMM algorithm stops before the
optimized portfolio reaches $5\%$. In order to match volatility $ \sigma
^{\star }>5\%$, we have to solve the strict $\sigma $-problem:
\begin{eqnarray}
w^{\star } &=&\arg \max \sum_{i=1}^{n}w_{i}\cdot \mu
_{i}-\sum_{i=1}^{n}\Delta w_{i}^{-}\left( c_{i}^{-}+\delta _{i}^{-}\Delta
w_{i}^{-}\right) -\sum_{i=1}^{n}\Delta w_{i}^{+}\left( c_{i}^{+}+\delta
_{i}^{+}\Delta w_{i}^{+}\right)   \label{eq:qcqp1} \\
&\mathrm{s.t.}&\left\{
\begin{array}{l}
\sum_{i=1}^{n}w_{i}+\sum_{i=1}^{n}\Delta w_{i}^{-}\left( c_{i}^{-}+\delta
_{i}^{-}\Delta w_{i}^{-}\right) +\sum_{i=1}^{n}\Delta w_{i}^{+}\left(
c_{i}^{+}+\delta _{i}^{+}\Delta w_{i}^{+}\right) =1 \\
w_{i}+\Delta w_{i}^{-}-\Delta w_{i}^{+}=\tilde{w}_{i} \\
\sqrt{w^{\top }\Sigma w}=\sigma ^{\star } \\
\mathbf{0}_{n}\leq w\leq \mathbf{1}_{n}%
\end{array}%
\right.   \notag
\end{eqnarray}%
This can be done by rewriting Problem (\ref{eq:qcqp1}) as a QCQP
program\footnote{See Appendix \ref{appendix:qcqp} on page
\pageref{appendix:qcqp}.}. In Figure \ref{fig:qc_mvo4}, we verify that QCQP
optimized portfolios such that $\sigma \left( w^{\star }\right) >5\%$ are in
fact not optimal, because they are dominated by portfolios with a higher
expected return and lower volatility.
\end{remark}

\section{Conclusion}

In this short note, we study mean-variance optimized portfolios with linear and
quadratic transaction costs. We show how the problem can be solved using the
techniques of quadratic programming and alternating direction method of
multipliers. We also illustrate how linear and quadratic transaction costs can
lead to different solutions and penalize the portfolio's return.
Moreover, the introduction of quadratic transaction costs opens a new field of research
when we consider transition management, asset ramp-up or portfolio scaling.


\appendix

\section*{Appendix}

\section{Mathematical results}

\subsection{Solution of the $y$-update in the case $\delta _{i}^{-}=\delta^{-}$ and
$\delta _{i}^{+}=\delta^{+}$} \label{appendix:y-update1}

We would like to solve the following nonlinear system of $3n+1$ equations:%
\begin{equation*}
\left\{
\begin{array}{l}
w_{i}-v_{i}+\lambda =0 \\
\Delta w_{i}^{-}-\Delta v_{i}^{-}+\lambda \left( c_{i}^{-}+2\delta
^{-}\Delta w_{i}^{-}\right) =0 \\
\Delta w_{i}^{+}-\Delta v_{i}^{+}+\lambda \left( c_{i}^{+}+2\delta
^{+}\Delta w_{i}^{+}\right) =0 \\
\sum_{i=1}^{n}\left( w_{i}+\Delta w_{i}^{-}\left( c_{i}^{-}+\delta^{-}\Delta
w_{i}^{-}\right) +\Delta w_{i}^{+}\left( c_{i}^{+}+\delta^{+}\Delta w_{i}^{+}\right) \right) =1%
\end{array}%
\right.
\end{equation*}%
The first $3n$ equations are equivalent to:%
\begin{equation*}
\left\{
\begin{array}{l}
w_{i}=v_{i}-\lambda  \\
\Delta w_{i}^{-}=\dfrac{\Delta v_{i}^{-}-\lambda c_{i}^{-}}{\lambda ^{-}} \\
\Delta w_{i}^{+}=\dfrac{\Delta v_{i}^{+}-\lambda c_{i}^{+}}{\lambda ^{+}}%
\end{array}%
\right.
\end{equation*}%
where $\lambda ^{-}=1+2\lambda \delta ^{-}$ and $\lambda ^{+}=1+2\lambda
\delta ^{+}$. The last equation then becomes:%
\begin{eqnarray*}
(\ast ) &\Leftrightarrow &\sum_{i=1}^{n}\left( v_{i}-\lambda \right)
+\sum_{i=1}^{n}\left( \frac{\Delta v_{i}^{-}-\lambda c_{i}^{-}}{\lambda ^{-}}%
\right) \left( c_{i}^{-}+\delta _{i}^{-}\frac{\Delta v_{i}^{-}-\lambda
c_{i}^{-}}{\lambda ^{-}}\right) + \\
&&\sum_{i=1}^{n}\left( \frac{\Delta v_{i}^{+}-\lambda c_{i}^{+}}{\lambda ^{+}%
}\right) \left( c_{i}^{+}+\delta _{i}^{+}\frac{\Delta v_{i}^{+}-\lambda
c_{i}^{+}}{\lambda ^{+}}\right) =1 \\
&\Leftrightarrow &\lambda ^{-^{2}}\lambda ^{+^{2}}\left(
\sum_{i=1}^{n}v_{i}-1\right) -\lambda \lambda ^{-^{2}}\lambda ^{+^{2}}n+ \\
&&\lambda ^{-}\lambda ^{+^{2}}\sum_{i=1}^{n}c_{i}^{-}\left( \Delta
v_{i}^{-}-\lambda c_{i}^{-}\right) +\lambda ^{+^{2}}\delta
^{-}\sum_{i=1}^{n}\left( \Delta v_{i}^{-}-\lambda c_{i}^{-}\right) ^{2}+ \\
&&\lambda ^{-^{2}}\lambda ^{+}\sum_{i=1}^{n}c_{i}^{+}\left( \Delta
v_{i}^{+}-\lambda c_{i}^{+}\right) +\lambda ^{-^{2}}\delta
^{+}\sum_{i=1}^{n}\left( \Delta v_{i}^{+}-\lambda c_{i}^{+}\right) ^{2}=0
\end{eqnarray*}%
We have $\lambda ^{-^{2}}=\left( 4\delta ^{-^{2}}\right) \lambda ^{2}+\left(
4\delta ^{-}\right) \lambda +1$, $\lambda ^{+^{2}}=\left( 4\delta
^{+^{2}}\right) \lambda ^{2}+\left( 4\delta ^{+}\right) \lambda +1$ and:
\begin{eqnarray*}
\lambda ^{-}\lambda ^{+^{2}} &=&\left( 8\delta ^{-}\delta ^{+^{2}}\right)
\lambda ^{3}+4\delta ^{+}\left( 2\delta ^{-}+\delta ^{+}\right) \lambda
^{2}+2\left( \delta ^{-}+2\delta ^{+}\right) \lambda +1 \\
\lambda ^{-^{2}}\lambda ^{+} &=&\left( 8\delta ^{-^{2}}\delta ^{+}\right)
\lambda ^{3}+4\delta ^{-}\left( \delta ^{-}+2\delta ^{+}\right) \lambda
^{2}+2\left( 2\delta ^{-}+\delta ^{+}\right) \lambda +1 \\
\lambda ^{-^{2}}\lambda ^{+^{2}} &=&\left( 16\delta ^{-^{2}}\delta
^{+^{2}}\right) \lambda ^{4}+16\delta ^{-}\delta ^{+}\left( \delta
^{-}+\delta ^{+}\right) \lambda ^{3}+ \\
&&4\left( \delta ^{-^{2}}+4\delta ^{-}\delta ^{+}+\delta ^{+^{2}}\right)
\lambda ^{2}+4\left( \delta ^{-}+\delta ^{+}\right) \lambda +1
\end{eqnarray*}%
We deduce that:%
\begin{eqnarray*}
(\ast ) &\Leftrightarrow &\left( \sum_{i=1}^{n}v_{i}-1\right) \left( \left(
16\delta ^{-^{2}}\delta ^{+^{2}}\right) \lambda ^{4}+16\delta ^{-}\delta
^{+}\left( \delta ^{-}+\delta ^{+}\right) \lambda ^{3}+4\left( \delta
^{-^{2}}+4\delta ^{-}\delta ^{+}+\delta ^{+^{2}}\right) \lambda ^{2}\right) +
\\
&&\left( \sum_{i=1}^{n}v_{i}-1\right) \left( 4\left( \delta ^{-}+\delta
^{+}\right) \lambda +1\right) -n\left( \left( 16\delta ^{-^{2}}\delta
^{+^{2}}\right) \lambda ^{5}+16\delta ^{-}\delta ^{+}\left( \delta
^{-}+\delta ^{+}\right) \lambda ^{4}\right) - \\
&&n\left( 4\left( \delta ^{-^{2}}+4\delta ^{-}\delta ^{+}+\delta
^{+^{2}}\right) \lambda ^{3}+4\left( \delta ^{-}+\delta ^{+}\right) \lambda
^{2}+\lambda \right) + \\
&&\left( \sum_{i=1}^{n}c_{i}^{-}\Delta v_{i}^{-}\right) \left( \left(
8\delta ^{-}\delta ^{+^{2}}\right) \lambda ^{3}+4\delta ^{+}\left( 2\delta
^{-}+\delta ^{+}\right) \lambda ^{2}+2\left( \delta ^{-}+2\delta ^{+}\right)
\lambda +1\right) - \\
&&\left( \sum_{i=1}^{n}c_{i}^{-^{2}}\right) \left( \left( 8\delta ^{-}\delta
^{+^{2}}\right) \lambda ^{4}+4\delta ^{+}\left( 2\delta ^{-}+\delta
^{+}\right) \lambda ^{3}+2\left( \delta ^{-}+2\delta ^{+}\right) \lambda
^{2}+\lambda \right) + \\
&&\left( \delta ^{-}\sum_{i=1}^{n}\Delta v_{i}^{-^{2}}\right) \left( \left(
4\delta ^{+^{2}}\right) \lambda ^{2}+\left( 4\delta ^{+}\right) \lambda
+1\right) - \\
&&\left( 2\delta ^{-}\sum_{i=1}^{n}\Delta v_{i}^{-}c_{i}^{-}\right) \left(
\left( 4\delta ^{+^{2}}\right) \lambda ^{3}+\left( 4\delta ^{+}\right)
\lambda ^{2}+\lambda \right) + \\
&&\left( \delta ^{-}\sum_{i=1}^{n}c_{i}^{-^{2}}\right) \left( \left( 4\delta
^{+^{2}}\right) \lambda ^{4}+\left( 4\delta ^{+}\right) \lambda ^{3}+\lambda
^{2}\right) + \\
&&\left( \sum_{i=1}^{n}c_{i}^{+}\Delta v_{i}^{+}\right) \left( \left(
8\delta ^{-^{2}}\delta ^{+}\right) \lambda ^{3}+4\delta ^{-}\left( \delta
^{-}+2\delta ^{+}\right) \lambda ^{2}+2\left( 2\delta ^{-}+\delta
^{+}\right) \lambda +1\right) - \\
&&\left( \sum_{i=1}^{n}c_{i}^{+^{2}}\right) \left( \left( 8\delta
^{-^{2}}\delta ^{+}\right) \lambda ^{4}+4\delta ^{-}\left( \delta
^{-}+2\delta ^{+}\right) \lambda ^{3}+2\left( 2\delta ^{-}+\delta
^{+}\right) \lambda ^{2}+\lambda \right) + \\
&&\left( \delta ^{+}\sum_{i=1}^{n}\Delta v_{i}^{+^{2}}\right) \left( \left(
4\delta ^{-^{2}}\right) \lambda ^{2}+\left( 4\delta ^{-}\right) \lambda
+1\right) - \\
&&\left( 2\delta ^{+}\sum_{i=1}^{n}\Delta v_{i}^{+}c_{i}^{+}\right) \left(
\left( 4\delta ^{-^{2}}\right) \lambda ^{3}+\left( 4\delta ^{-}\right)
\lambda ^{2}+\lambda \right) + \\
&&\left( \delta ^{+}\sum_{i=1}^{n}c_{i}^{+^{2}}\right) \left( \left( 4\delta
^{-^{2}}\right) \lambda ^{4}+\left( 4\delta ^{-}\right) \lambda ^{3}+\lambda
^{2}\right)  \\
&=&0
\end{eqnarray*}%
We obtain a quintic equation:%
\begin{equation*}
\alpha _{5}\lambda ^{5}+\alpha _{4}\lambda ^{4}+\alpha _{3}\lambda
^{3}+\alpha _{2}\lambda ^{2}+\alpha _{1}\lambda +\alpha _{0}=0
\end{equation*}%
where:%
\begin{equation*}
\alpha _{5}=16n\delta ^{-^{2}}\delta ^{+^{2}}
\end{equation*}%
\begin{eqnarray*}
\alpha _{4} &=&\left( 16\delta ^{-^{2}}\delta ^{+^{2}}\right) \left(
\sum_{i=1}^{n}v_{i}-1\right) -16n\delta ^{-}\delta ^{+}\left( \delta
^{-}+\delta ^{+}\right) - \\
&&4\delta ^{-}\delta ^{+}\left( \delta
^{+}\sum_{i=1}^{n}c_{i}^{-^{2}}+\delta
^{-}\sum_{i=1}^{n}c_{i}^{+^{2}}\right)
\end{eqnarray*}%
\begin{eqnarray*}
\alpha _{3} &=&16\delta ^{-}\delta ^{+}\left( \delta ^{-}+\delta ^{+}\right)
\left( \sum_{i=1}^{n}v_{i}-1\right) -4n\left( \delta ^{-^{2}}+4\delta
^{-}\delta ^{+}+\delta ^{+^{2}}\right) - \\
&&4\left( \delta ^{-}+\delta ^{+}\right) \left( \delta
^{+}\sum_{i=1}^{n}c_{i}^{-^{2}}+\delta
^{-}\sum_{i=1}^{n}c_{i}^{+^{2}}\right)
\end{eqnarray*}%
\begin{eqnarray*}
\alpha _{2} &=&4\left( \delta ^{-^{2}}+4\delta ^{-}\delta ^{+}+\delta
^{+^{2}}\right) \left( \sum_{i=1}^{n}v_{i}-1\right) -4n\left( \delta
^{-}+\delta ^{+}\right) + \\
&&4\left( \delta ^{+^{2}}\sum_{i=1}^{n}c_{i}^{-}\Delta v_{i}^{-}+\delta
^{-^{2}}\sum_{i=1}^{n}c_{i}^{+}\Delta v_{i}^{+}\right) -\left( \delta
^{-}+4\delta ^{+}\right) \sum_{i=1}^{n}c_{i}^{-^{2}}- \\
&&\left( 4\delta ^{-}+\delta ^{+}\right) \sum_{i=1}^{n}c_{i}^{+^{2}}+4\delta
^{-}\delta ^{+}\left( \delta ^{+}\sum_{i=1}^{n}\Delta v_{i}^{-^{2}}+\delta
^{-}\sum_{i=1}^{n}\Delta v_{i}^{+^{2}}\right)
\end{eqnarray*}%
\begin{eqnarray*}
\alpha _{1} &=&4\left( \delta ^{-}+\delta ^{+}\right) \left(
\sum_{i=1}^{n}v_{i}-1\right) -n+4\left( \delta
^{+}\sum_{i=1}^{n}c_{i}^{-}\Delta v_{i}^{-}+\delta
^{-}\sum_{i=1}^{n}c_{i}^{+}\Delta v_{i}^{+}\right) - \\
&&\left( \sum_{i=1}^{n}c_{i}^{-^{2}}+\sum_{i=1}^{n}c_{i}^{+^{2}}\right)
+4\delta ^{-}\delta ^{+}\left( \sum_{i=1}^{n}\Delta
v_{i}^{-^{2}}+\sum_{i=1}^{n}\Delta v_{i}^{+^{2}}\right)
\end{eqnarray*}%
and:%
\begin{equation*}
\alpha _{0}=\left( \sum_{i=1}^{n}v_{i}-1\right) +\left(
\sum_{i=1}^{n}c_{i}^{-}\Delta v_{i}^{-}+\sum_{i=1}^{n}c_{i}^{+}\Delta
v_{i}^{+}\right) +\left( \delta ^{-}\sum_{i=1}^{n}\Delta
v_{i}^{-^{2}}+\delta ^{+}\sum_{i=1}^{n}\Delta v_{i}^{+^{2}}\right)
\end{equation*}

\subsection{Solution of the $y$-update in the case $\delta_{i}^{-} \neq \delta_{j}^{-}$ and
$\delta_{i}^{+} \neq \delta_{j}^{+}$}
\label{appendix:y-update2}

We consider the following optimization problem:%
\begin{eqnarray*}
y^{\left( k+1\right) } &=&\arg \min \frac{1}{2}\left( y-v_{y}^{\left(
k+1\right) }\right) ^{\top }\left( y-v_{y}^{\left( k+1\right) }\right)  \\
&\mathrm{s.t.}&A_{1}y+y^{\top }C_{1}y-B_{1}=0
\end{eqnarray*}%
Following Park and Boyd (2017), the Lagrangian is given by:%
\begin{eqnarray*}
\mathcal{L}\left( y,\lambda \right)  &=&\frac{1}{2}\left( y-v_{y}^{\left(
k+1\right) }\right) ^{\top }\left( y-v_{y}^{\left( k+1\right) }\right)
+\lambda \left( A_{1}y+y^{\top }C_{1}y-B_{1}\right)  \\
&=&\frac{1}{2}y^{\top }\left( I_{3n}+2\lambda C_{1}\right) y+\left( \lambda
A_{1}-v_{y}^{\left( k+1\right) \top }\right) y+ \\
&&\left( \frac{1}{2}v_{y}^{\left( k+1\right) \top }v_{y}^{\left( k+1\right)
}-\lambda B_{1}\right)
\end{eqnarray*}%
The first order conditions are:%
\begin{equation*}
\left\{
\begin{array}{l}
\left( I_{3n}+2\lambda C_{1}\right) y+\lambda A_{1}^{\top }-v_{y}^{\left(
k+1\right) }=\mathbf{0}_{3n} \\
A_{1}y+y^{\top }C_{1}y-B_{1}=0%
\end{array}%
\right.
\end{equation*}%
Therefore, we have:%
\begin{equation*}
y=\left( I_{3n}+2\lambda C_{1}\right) ^{-1}\left( v_{y}^{\left( k+1\right)
}-\lambda A_{1}^{\top }\right)
\end{equation*}%
It follows that the equality constraint becomes:%
\begin{eqnarray*}
A_{1}\left( I_{3n}+2\lambda C_{1}\right) ^{-1}\left( v_{y}^{\left(
k+1\right) }-\lambda A_{1}^{\top }\right) + && \\
\left( v_{y}^{\left( k+1\right) }-\lambda A_{1}^{\top }\right) ^{\top
}\left( I_{3n}+2\lambda C_{1}\right) ^{-1}C_{1}\left( I_{3n}+2\lambda
C_{1}\right) ^{-1}\left( v_{y}^{\left( k+1\right) }-\lambda A_{1}^{\top
}\right) -B_{1} &=&0
\end{eqnarray*}%
Since $C_{1}$ is a diagonal matrix, $\left( I_{3n}+2\lambda C_{1}\right)
^{-1}$ is also a diagonal matrix. It follows that the previous equation is
equivalent to:%
\begin{equation*}
\sum_{i=1}^{3n}\frac{A_{1,i}\left( v_{y,i}^{\left( k+1\right) }-\lambda
A_{1,i}\right) }{1+2\lambda \left( C_{1}\right) _{i,i}}+\sum_{i=1}^{3n}\frac{%
\left( C_{1}\right) _{i,i}\left( v_{y,i}^{\left( k+1\right) }-\lambda
A_{1,i}\right) ^{2}}{\left( 1+2\lambda \left( C_{1}\right) _{i,i}\right) ^{2}%
}-B_{1}=0
\end{equation*}%
By replacing $A_{1,i}$, $B_{1}$, $\left( C_{1}\right) _{i,i}$ and $%
v_{y,i}^{\left( k+1\right) }$ by their values, we obtain the following
nonlinear equation:%
\begin{eqnarray*}
\sum_{i=1}^{n}\left( v_{i}-\lambda \right) +\sum_{i=1}^{n}\frac{%
c_{i}^{-}\left( \Delta v_{i}^{-}-\lambda c_{i}^{-}\right) }{1+2\lambda
\delta _{i}^{-}}+\sum_{i=1}^{n}\frac{c_{i}^{+}\left( \Delta
v_{i}^{+}-\lambda c_{i}^{+}\right) }{1+2\lambda \delta _{i}^{+}}+ && \\
\sum_{i=1}^{n}\frac{\delta _{i}^{-}\left( \Delta v_{i}^{-}-\lambda
c_{i}^{-}\right) ^{2}}{\left( 1+2\lambda \delta _{i}^{-}\right) ^{2}}%
+\sum_{i=1}^{n}\frac{\delta _{i}^{+}\left( \Delta v_{i}^{+}-\lambda
c_{i}^{+}\right) ^{2}}{\left( 1+2\lambda \delta _{i}^{+}\right) ^{2}}-1 &=&0
\end{eqnarray*}%
Park and Boyd (2017) noticed that the derivative of the lefthand side is negative, meaning that the
function is decreasing and has a unique root. They then suggested to solve this equation using the bisection
method. Once the optimal value $\lambda ^{\star }$ is found, the solution $%
y^{\left( k+1\right) }$ is given by:%
\begin{equation*}
\left\{
\begin{array}{l}
w_{i}=v_{i}-\lambda ^{\star } \\
\Delta w_{i}^{-}=\dfrac{\Delta v_{i}^{-}-\lambda ^{\star }c_{i}^{-}}{%
1+2\lambda ^{\star }\delta _{i}^{-}} \\
\Delta w_{i}^{+}=\dfrac{\Delta v_{i}^{+}-\lambda ^{\star }c_{i}^{+}}{%
1+2\lambda ^{\star }\delta _{i}^{+}}%
\end{array}%
\right.
\end{equation*}

\subsection{QCQP formulation of the strict $\sigma $-problem}
\label{appendix:qcqp}

In the case of the strict $\sigma $-problem, the optimization problem becomes:
\begin{eqnarray*}
x^{\star } &=&\arg \min -x^{\top }R+x^{\top }Qx \\
&\mathrm{s.t.}&\left\{
\begin{array}{l}
A_{1}x+x^{\top }C_{1}x=B_{1} \\
A_{2}x=B_{2} \\
x^{\top }C_{3}x=B_{3} \\
x^{-}\leq x\leq x^{+}%
\end{array}%
\right.
\end{eqnarray*}%
where $x=\left( w,\Delta w^{-},\Delta w^{+}\right) $,
\begin{equation*}
Q=\left(
\begin{array}{ccc}
\mathbf{0}_{n,n} & \mathbf{0}_{n,n} & \mathbf{0}_{n,n} \\
\mathbf{0}_{n,n} & \Delta ^{-} & \mathbf{0}_{n,n} \\
\mathbf{0}_{n,n} & \mathbf{0}_{n,n} & \Delta ^{+}%
\end{array}%
\right)
\end{equation*}%
and:%
\begin{equation*}
R=\left(
\begin{array}{c}
\mu  \\
-c^{-} \\
-c^{+}%
\end{array}%
\right)
\end{equation*}%
For the equality constraints, we obtain:
\begin{equation*}
\left(
\begin{array}{c}
A_{1} \\
A_{2}%
\end{array}%
\right) =\left(
\begin{array}{ccc}
\mathbf{1}_{n}^{\top } & \left( c^{-}\right) ^{\top } & \left( c^{+}\right)
^{\top } \\
I_{n} & I_{n} & -I_{n}%
\end{array}%
\right)
\end{equation*}%
and:%
\begin{equation*}
\left(
\begin{array}{c}
B_{1} \\
B_{2} \\
B_{3}%
\end{array}%
\right) =\left(
\begin{array}{c}
1 \\
\tilde{w} \\
\sigma ^{\star 2}%
\end{array}%
\right)
\end{equation*}%
where $\sigma ^{\star }$ is the targeted volatility of the portfolio. The
matrices $C_{1}$ and $C_{3}$ are defined as follows:%
\begin{equation*}
C_{1}=\left(
\begin{array}{ccc}
\mathbf{0}_{n,n} & \mathbf{0}_{n,n} & \mathbf{0}_{n,n} \\
\mathbf{0}_{n,n} & \Delta ^{-} & \mathbf{0}_{n,n} \\
\mathbf{0}_{n,n} & \mathbf{0}_{n,n} & \Delta ^{+}%
\end{array}%
\right)
\end{equation*}%
and:%
\begin{equation*}
C_{3}=\left(
\begin{array}{ccc}
\Sigma  & \mathbf{0}_{n,n} & \mathbf{0}_{n,n} \\
\mathbf{0}_{n,n} & \mathbf{0}_{n,n} & \mathbf{0}_{n,n} \\
\mathbf{0}_{n,n} & \mathbf{0}_{n,n} & \mathbf{0}_{n,n}%
\end{array}%
\right)
\end{equation*}%
The bounds remain the same: $x^{-}=\mathbf{0}_{3n}$ and:
\begin{equation*}
x^{+}=\left(
\begin{array}{c}
\mathbf{1}_{n} \\
\tilde{w} \\
\mathbf{1}_{n}-\tilde{w}%
\end{array}%
\right)
\end{equation*}%
Again, the optimal solution $w^{\star }$ is given by the following
relationship:
\begin{equation*}
w^{\star }=\left(
\begin{array}{ccc}
I_{n} & \mathbf{0}_{n,n} & \mathbf{0}_{n,n}%
\end{array}%
\right) x^{\star }
\end{equation*}


\begin{thebibliography}{99}

\bibitem{} \textsc{Boyd}, S., \textsc{Parikh}, N., \textsc{Chu}, E.,
    \textsc{Peleato}, B., and \textsc{Eckstein}, J. (2010),
    Distributed Optimization and Statistical Learning via the
    Alternating Direction Method of Multipliers, \textit{Foundations and Trends\textregistered\ in
    Machine learning}, 3(1), pp. 1-122.

\bibitem{} \textsc{Gabay}, D., and \textsc{Mercier}, B. (1976), A Dual
    Algorithm for the Solution of Nonlinear Variational Problems via Finite Element
    Approximation, \textit{Computers \& Mathematics with Applications}, 2(1),
    pp. 17-40.

\bibitem{} \textsc{Lecesne}, L., and \textsc{Roncoroni}, A. (2019a), Optimal
    Allocation in the S\&P 600 Under Size-driven Illiquidity, \textit{ESSEC
    Working Paper}.

\bibitem{} \textsc{Lecesne}, L., and \textsc{Roncoroni}, A. (2019b), How Should
    Funds Decisions and Performances React to Size-Driven Liquidity Friction,
    \textit{ESSEC Working Paper}.

\bibitem{} \textsc{Markowitz}, H. (1952), Portfolio Selection,
    \textit{Journal of Finance}, 7(1), pp. 77-91.

\bibitem{} \textsc{Park}, J., and \textsc{Boyd}, S. (2017),
General Heuristics for Nonconvex Quadratically Constrained Quadratic Programming,
\textit{arXiv}, 1703.07870.


\bibitem{} \textsc{Perrin}, S., and \textsc{Roncalli}, T. (2019), Machine
    Learning Algorithms and Portfolio Optimization, in Jurczenko, E. (Ed.),
    \textit{Machine Learning in Asset Management}, ISTE Press -- Elsevier,
    forthcoming.

\bibitem{} \textsc{Roncalli}, T. (2013), \textit{Introduction to Risk Parity
    and Budgeting}, Chapman and Hall/CRC Financial Mathematics Series.

\bibitem{} \textsc{Scherer}, B. (2007), \textit{Portfolio Construction \& Risk
    Budgeting}, Third edition, Risk Books.
\end{thebibliography}
\end{document}